\documentclass[prb,aps,amsmath,amssymb,showpacs,floatfix,twocolumn]{revtex4}
\usepackage{graphicx,epsfig}
\usepackage[usenames]{color}

\usepackage{soul}

\newcommand{\be}{\begin{equation}}
\newcommand{\ee}{\end{equation}}
\newcommand{\bea}{\begin{eqnarray}}
\newcommand{\eea}{\end{eqnarray}}

\begin{document}
\title{Non-equilibrium transport theory of the singlet-triplet
transition: perturbative approach} 

\author{B. Horv\'ath$^1$, B Lazarovits$^{1,2}$, and G. Zar\'and$^1$}
\affiliation{
$^1$Theoretical Physics Department, Institute of Physics, Budapest University of 
Technology and Economics, Budafoki \'ut 8, H-1521 Hungary\\
$^2$Research Institute for Solid State Physics and Optics of the Hungarian Academy 
of Sciences,
Konkoly-Thege M. \'ut 29-33., H-1121 Budapest, Hungary}

\begin{abstract}
We use a simple iterative perturbation theory to study the singlet-triplet 
(ST) transition in 
lateral and vertical quantum dots, modeled by 
the non-equilibrium two-level  
Anderson model. To a great
surprise, the region of stable perturbation theory extends to 
relatively strong interactions, and 
 this simple approach is able to reproduce all
experimentally-observed features of the ST transition, including 
the formation of a dip in the differential conductance of a lateral dot
indicative of the 
two-stage Kondo effect, or the maximum in the linear conductance
around the transition point. Choosing the right starting point to the
perturbation theory is, however, crucial to obtain reliable and
meaningful results.
\end{abstract}

\maketitle

\section{Introduction}

With the evolution of mesoscopic and nano-scale technology and 
electron lithography, physicists gain more and more control over 
sub-micron structures and molecular 
devices.\cite{hp1,hp2,hp3,hp4} By exploiting 
cutting-edge 
semiconductor technology, experimentalists can now not only 
build artificial atoms and molecules to trap and manipulate 
 individual electrons and their spin,\cite{rev1,rev2}
but they are also able contact and gate real molecules,\cite{Heath} 
and build simple  molecular devices out of them such as 
the single electron transistor. 
These devices are promising candidates for our 
future electronics, and might possibly be used as building blocks 
of spintronics devices and spin-based quantum computers, too.\cite{Loss}   
Moreover, the unprecedented control of these minuscule structures opens
new and fascinating possibilities to build and study hybrid structures,
\cite{Martinek,Calvo,Belzig,Kouwenhoven,Hewson} entangle electron spins,\cite{Szabolcs} 
and also to realize and study simple quantum 
systems in the close vicinity of quantum phase 
transitions under non-equlibrium 
conditions.\cite{Goldhaber,Kogan,Vojta,Zarand_2imp,Affleck2imp}

Understanding the physical properties of these tiny electronic circuits 
represents a major challenge to theoretical as well as to experimental 
physicists.  Due to their extremely small size, 
electron-electron interaction is often dominant in these structures, 
and moreover, as mentioned before, they are operated under non-equilibrium
conditions. Although theoretical physicists devoted a lot of effort to  
design methods that are able to tackle these difficulties, so far 
none of the proposed methods proved to be entirely 
successful in describing strongly interacting non-equilibrium
systems: Monte Carlo methods are presently unable to reach the required 
precision,\cite{MC_Egger,Han} Bethe Ansatz methods can be used for
a few models only, and are still in an experimental stage,\cite{Andrei,Andrei_err,Saleur}  
and perturbative renormalization group methods can only 
reach a particular region of 
the parameter space.\cite{Rosch,Schoeller,Kehrein,Meden} Maybe numerical
renormalization group methods are currently the most reliable
techniques to study these non-equilibrium
 systems,\cite{Scholl,Anders,Saleur} however, they scale
 very badly with the number of states involved, and to compute the
 transport through just two levels in the presence of
 interaction seems to be numerically too demanding.    

In view of the above-described situation, perturbative 
methods such as iterative perturbation theory 
(IPT),\cite{Yeyati,Kotliar,Aligia} e.g., are of great value, 
especially, since many 
well-contacted or almost open systems are {\em de facto} in the
perturbative regime.  For this reason, several molecular electronics 
groups combine the GW method with standard \emph{ab initio} tools to compute
transport through molecules.\cite{gw} Unfortunately, 
however, very little is known about the  reliability of this 
approach. It is, e.g., rather easy to obtain 
 spontaneous local symmetry breaking (e.g., magnetic moment formation), and
thereby obtain physically incorrect results.\cite{BerciPRB} 

In the present paper we investigate systematically the perturbative 
approach for the simplest possible non-trivial molecular system, a 
two-level Anderson model with two degenerate or nearly degenerate
electron levels. This simple model describes many experimental
systems, and it displays  rich physical properties. 
Having an odd number of electrons on the two levels, e.g., one can recover 
an almost  SU(4) symmetrical Kondo  state.\cite{Zarandsu4,Bordasu4,Sasakisu4,Herrerosu4,Choisu4}
In case of two electrons, on the other hand, 
one finds a transition
(cross-over) between a state, where the two electrons 
are bound into a \emph{singlet}, and another state, where the two
electrons'
spin is aligned into a \emph{triplet}, which is then completely or
partially screened by the spin of the conduction electrons in 
the leads.\cite{Vojta,Pustilnik,Zarand,Kogan,Wiel,Granger}
Here we focus our attention to this second,  most
interesting regime, also nicknamed as the singlet-triplet (ST)
transition. We focus on the most realistic case,\cite{Pustilnik} where two
independent conduction electron channels are coupled to the dot 
levels, and therefore the singlet-triplet transition is just a
cross-over.\cite{Wiel,Granger,Pustilnik,Zarand} 
This must be contrasted to the case of a single coupled 
conduction electron  channel, when the transition is a true
Kosterlitz-Thouless quantum phase
transition.\cite{Vojta,Kogan} Non-equilibrium transport in this 
latter case has been investigated recently using the so-called
non-crossing approximation (NCA).\cite{aligia_nca} 

Describing the case of two coupled channels under  non-equilibrium 
conditions represents a theoretical challange.
The deep singlet side of the transition is accessible 
through perturbative renormalization  group methods, which were used
successfully to analyze the non-equilibrium transport in this
perturbative regime.\cite{paaske} Here we focus on the transition, and 
investigate, whether a simple perturbative approach 
is able to account for the transition between the two correlated states. 
To our surprise,
 perturbation theory captures  all qualitative
 features of the transition, including the two-stage Kondo 
effect, or the conductance maximum around the transition 
point.\cite{Pustilnik,Zarand,Wiel,Granger} 
However, to obtain accurate perturbative 
results, one  needs to expand around the best possible non-interacting
theory.\cite{Kotliar,Yeyati,Aligia}  
This we achieve by introducing self-consistently
determined countertems, which is essentially equivalent 
to doing perturbation theory around the Hartree 
solution.

The simple perturbative calculation is, however, unable to 
capture correctly the Kondo scales, i.e. the energy scales 
below which the electron spins on the dot are 
screened or bound to a singlet. As we show in 
a Ref.~\onlinecite{future}, this
shortcoming of the  iterative perturbation theory 
can be overcome to a large extent by the so-called 
Fluctuation Exchange Approximation (FLEX) extended 
to non-equilibrium  situations.  

The paper is organized as follows. In Secs.~\ref{modelsec} 
and \ref{iptsec},
we introduce the non-equilibrium two-level Anderson model, 
and describe the 
iterative perturbation theory  
used to solve the non-equilibrium Anderson model. 
In Sec.~\ref{transport} we show how to obtain the transport 
properties.  Some computational details  and the 
limitations of the iterative perturbation scheme are given in 
sections~\ref{compsec} and \ref{numerics}. 
In Sec.~\ref{sym}, we discuss the results obtained 
for completely symmetrical quantum dots with equal 
level widths, while in Sec.~\ref{assymsec}  
 results for dots with more generic parameters are presented. 
Our conclusions are summarized  conclude in Sec.~\ref{conclu}, 
 and  some technical details are given in Appendix~\ref{firstapp}.

\section{Theoretical framework}

\subsection{Model}\label{modelsec}

The physical properties of a quantum dot with even number of electrons are
usually well-captured by a simple two-level 
Anderson Hamiltonian.
This can be  written as a sum  of a non-interacting and an
interacting term:  $H = H_{0}+H_{\rm int}$. The non-interacting part, 
$H_0$, can be further divided onto three terms, 
\be
 H_{0}=H_{\rm cond}+H_{\rm hyb}+H_{0,\rm dot}. 
\ee 
Here $H_{\rm cond}$ describes conduction electrons in the
leads, $H_{\rm hyb}$ accounts for the hybridization
between the conduction electrons and the electrons residing on the
dot,  and $H_{0,\rm dot}$ stands for the non-interacting part of the dot 
Hamiltonian,
\begin{equation}
H_{0,\rm dot} = \sum_{i,\sigma}\varepsilon_{i}d_{i\sigma}^{\dagger}d_{i\sigma}\;,
\end{equation}
with $d_{i\sigma}^{\dagger}$ the creation operator 
of a dot electron of spin $\sigma$  on dot 
level $i$ of energy $\varepsilon_{i}$ ($i\in(+,-)$).
  
The precise form of $H_{\rm cond}$ and $H_{\rm hyb}$ 
depends on the geometrical arrangement of the quantum dot.  
 In case of a 
lateral quantum dot close to pinch-off, there is only a single
 conduction electron mode coupled to the dot, and correspondingly, 
\begin{eqnarray}
   H^{\rm lat}_{\rm cond} & = & \sum_{\xi,\alpha,\sigma}
\xi_{\alpha}c_{\xi\alpha\sigma}^{\dagger}c_{\xi\alpha\sigma}\;,\\
   H^{\rm lat}_{\rm hyb} & = & \sum_{\alpha,i,\xi,\sigma}t_{\alpha
     i}(c_{\xi\alpha\sigma}^{\dagger}d_{i\sigma}+h.c.)\;. 
\end{eqnarray}
Here $c_{\xi\alpha\sigma}^{\dagger}$ creates a conduction electron in
the left or right lead  ($\alpha\in(L,R)$) with energy  
$\xi_{\alpha}=\xi+\mu_{\alpha}$, with 
$\mu_{\alpha}=eV_{\alpha}$  the bias applied on lead $\alpha$. 
The $t_{\alpha i}$'s denote  tunneling matrix
elements between lead $\alpha$ and dot level $i$, and  
for the sake of simplicity here we assume them to be   
spin- and energy independent. 

Vertical quantum dots\cite{sasaki,Pust2,pusgla} or
nanotubes\cite{nanotubes} 
are, on the other hand,  often
connected to leads with a large surface and therefore in each lead
there are typically many different conduction electron
modes which couple to a given dot state. However, for each dot state
$i$ and lead $\alpha$, one can construct a single linear combination 
of modes, $d_{i\sigma}^\dagger\to c_{\xi,i\alpha\sigma}^{\dagger}$, which
hybridizes with $d_{i\sigma}$. In many cases, one can
assume (based on symmetry considerations or in view of the chaotic
shape of the dot wave functions and the large number of lead modes) 
that $c_{\xi,i\; \alpha\sigma}^{\dagger}$ are independent of
each other for $i=\pm$ i.e. $\{c_{\xi,+,\alpha,\sigma}^{\dagger},c_{\xi,-,\alpha,\sigma}\}=0$. 
Under this assumption,  $H_{\rm cond}$ and $ H_{\rm hyb}$ assume the form, 
\begin{eqnarray}
   H^{\rm vert}_{\rm cond} & = & \sum_{\xi,i,\alpha,\sigma}\xi_{\alpha}c_{\xi i\alpha\sigma}^{\dagger}c_{\xi i\alpha\sigma}\;,\\
   H^{\rm vert}_{\rm hyb} & = & \sum_{\alpha,i,\xi,\sigma}
t_{\alpha i}(c_{\xi i\alpha\sigma}^{\dagger}d_{i\sigma}+h.c.)\;.
\end{eqnarray}
Notice that the non-equilibrium condition is taken into account through
a lead-dependent  shift of the chemical potentials, and  
the interaction between the 
conduction electrons is neglected. 

In general, the hybridization matrix elements $t_{\alpha i}$ are
independent of each-other. However, for simplicity, in 
this paper we focus on completely {\em symmetrical} quantum dots
and assume further that one of the dot 
levels is even ($+$), while the other is odd under reflection ($-$). 
As a consequence, for a lateral dot the tunneling matrix elements 
have a simple structure: $t_{L,+}=t_{R,+}$  and
$t_{L,-}=-t_{R,-}$. Although this choice may seem
to be too special, this symmetrical model still captures
most of the generic features of a two-level quantum 
dot.\cite{Pustilnik,Zarand}
 
In our model, electron-electron interaction on the dot is taken into 
account by the term:
\begin{equation}
 H_{\rm int} =
 \frac{U}{2}\left(\sum_{i,\sigma}n_{i\sigma}-2\right)^{2}-J\;\vec{S}^{2}\;,
\label{intham}
\end{equation}
where $n_{i\sigma}=d_{i\sigma}^{\dagger}d_{i\sigma}$ and 
$\vec{S}=\frac{1}{2}\sum\limits_{i,\sigma,\sigma'}d_{i\sigma}^{\dagger}
\vec{\sigma}_{\sigma\sigma'}d_{i\sigma'}$ with $\vec{\sigma}$ denoting
the Pauli matrices,  $\vec{\sigma}=(\sigma_x,\sigma_y,\sigma_z)$.
Here the coupling constant $U$ denotes the on-site Coulomb
interaction, and accounts  for the charging energy of the 
dot, while $J$ stands for the Hund's rule coupling,
favoring a ferromagnetic  alignment of the dot electron spins $(J>0)$.
In this work,  we focus our attention to  the regime where approximately 
two electrons reside on the dot. This assumption is implicit in the 
way we expressed the Coulomb term in Eq.~\eqref {intham}, which obviously 
possesses a particle-hole symmetry discussed later.

Although  Eq.~\eqref{intham} provides a rather symmetrical and 
compact expression for the interaction term, for the purpose of a
systematic perturbation theory,  it is useful to rewrite it in a
normal ordered form, 
\begin{equation}
 H_{\rm int}= :H_{\rm int}: -\left(\frac {3U} 2 +    
\frac {3J} 4\right)  \sum_{i,\sigma}n_{i\sigma} \;, 
\label{:intham:}
\end{equation}
and treat only the normal ordered part as a perturbation, 
$:H_{\rm int}:$, while adding  the second term in this expression 
to $H_{0,\rm dot}$, 
\be 
 \epsilon_{i}\to E_{i}\equiv
\epsilon_{i}-\frac{3J}{4} - \frac {3U} 2 \;. 
\label{levels}
\ee 
We remark that for a two-level dot close to half filling, $\epsilon_{i}\approx 0$, and the point $\epsilon_{+}= 
\epsilon_{-}=0$ corresponds to the case where the two levels are
exactly degenerate, and there are exactly two electrons on the dot ($\langle
\sum_{i\sigma}n_{i\sigma}\rangle=2$). 

\begin{figure}[t]
\hskip3cm\includegraphics[width=240pt,clip=true]{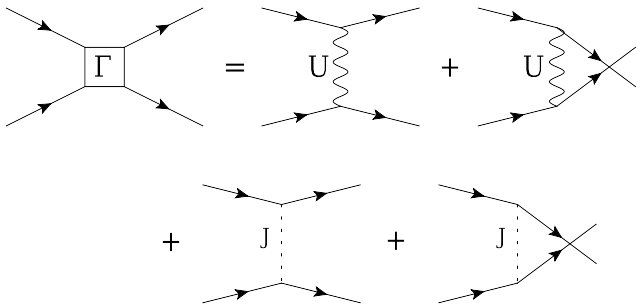}
\caption{The antisymmetrized vertex function containing the Coulomb interaction and 
the Hund's rule coupling.}\label{vertex}
\end{figure}

\begin{figure}[b]
\begin{center}
\includegraphics[width=5.5cm,clip=true]{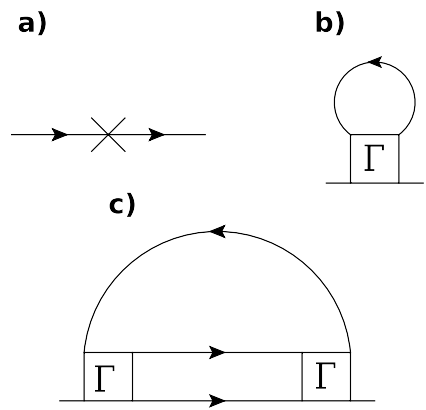}
\end{center}
\caption{
Counterterm ($a$),  first order ($b$) and 
second order  ($c$) contributions to the self-energy. The lines denote
non-interacting Keldysh Green's functions.}
\label{self}
\end{figure}

For later purposes, let us give here the explicit form of $:H_{\rm
  int}:$ as obtained  by exploiting fermionic  
commutation relations,
\begin{eqnarray}
 :H_{\rm int}: & = &
-\frac{J}{2}\sum_{\sigma}\sum_{i\neq j}d_{i\sigma}^{\dagger}d_{j\bar{\sigma}}^{\dagger}d_{j\sigma}d_{i\bar{\sigma}}+\nonumber\\
& + & \left(\frac{U}{2}+\frac{3J}{4}\right)
\sum_{i,\sigma}d_{i\sigma}^{\dagger}d_{i\bar{\sigma}}^{\dagger}d_{i\bar{\sigma}}d_{i\sigma}+\nonumber\\
& + &
\left(\frac{U}{2}-\frac{J}{4}\right)\sum_{i\neq j,\sigma}d_{i\sigma}^{\dagger}d_{j\sigma}^{\dagger}d_{j\sigma}d_{i\sigma}
+\nonumber\\
& + &\left(\frac{U}{2}+\frac{J}{4}\right)
\sum_{i\neq j,\sigma}d_{i\sigma}^{\dagger}d_{j\bar{\sigma}}^{\dagger}d_{j\bar{\sigma}}d_{i\sigma}\;,\label{newintham}
\end{eqnarray}
where $\bar{\sigma}=-\sigma$. To simplify perturbation theory, 
it is useful to  collect all fourth order terms and rewrite $:H_{\rm int}:$
in terms of the antisymmetrized  interaction vertex shown in Fig.~\ref{vertex},
$\Gamma_{i\sigma\;n\tilde\sigma}^{j\sigma'\;m\tilde\sigma'}$, as
\begin{equation}
 :{H}_{\rm int}:=\sum_{\substack{i,j,m,n \\ \sigma,\sigma',\tilde{\sigma},\tilde{\sigma}'}}
\frac{1}{4}\;\Gamma_{i\sigma\;n\tilde\sigma}^{j\sigma'\;m\tilde\sigma'}
d_{j\sigma'}^{\dagger}d_{m\tilde{\sigma}'}^{\dagger}d_{n\tilde{\sigma}}d_{i\sigma}\;. 
\label{eq:vert_ham}
\end{equation}
Using the fact that 
$\Gamma_{i\sigma\;n\tilde\sigma}^{j\sigma'\;m\tilde\sigma'}$ is 
anti-symmetrical under exchanging $i\sigma\leftrightarrow n\tilde\sigma$
and $j\sigma'\leftrightarrow m\tilde\sigma'$, it is easy to read out
the non-vanishing matrix elements of 
$\Gamma_{i\sigma\;n\tilde\sigma}^{j\sigma'\;m\tilde\sigma'}$ from
Eq.~\eqref{newintham}. Although the vertex 
$\Gamma$ has formally 256 elements, 
as a consequence of spin and level-conservation in Eq. (\ref{newintham}), 
only thirty-two of these enter actual calculations.
The advantage of the symmetrical form, Eq.~\eqref{eq:vert_ham}, is
that it allows one to treat the Coulomb interaction and the 
 Hund's rule coupling on equal footing:
one can formally (and numerically) evaluate the self energy 
diagrams shown schematically in Fig.~\ref{self}, with the specific
form of the interaction hidden  in  the internal structure of these diagrams showing in Fig.~\ref{vertex}.

Before proceeding with the presentation of  our non-equilibrium perturbation 
theory, let us shortly discuss the symmetry properties of the
Hamiltonian. Clearly, $H$ is invariant under SU(2)
{\em spin rotations}. This implies that, in the absence of an external
magnetic field, all
Green's functions are diagonal in the spin labels and also
independent of them. 

In addition to spin SU(2) symmetry, depending on the specific Hamiltonian 
parameters, $H$ may also possess a discrete {\em electron-hole} symmetry. 
In fact, throughout the present paper 
we shall restrict ourself to the simple case, where the two levels 
$\epsilon_\pm$ are shifted symmetrically away from 0,
\be 
\epsilon_\pm =  \pm \Delta/2\;.
\ee
Then, for $|t_{\alpha +}| = |t_{\alpha -}|$, the Hamiltonian 
is also invariant under the following transformation:
\bea 
(d^\dagger_{+,\uparrow},\;d^\dagger_{+,\downarrow}) &\leftrightarrow &
(d_{-,\downarrow},\;-\;d_{-,\uparrow})\;,
\nonumber
\\
(c^\dagger_{\xi, L \uparrow},\;c^\dagger_{\xi,L\downarrow}) &\leftrightarrow &
(-\;c_{-\xi, R\downarrow},\;c_{-\xi,R\uparrow})\;.
\label{eh1}
\eea
This symmetry implies that the occupation 
of the dot is exactly $\langle \sum_{i,\sigma}n_{i\sigma}\rangle\equiv2$ for any 
interaction strength and splitting $\Delta$, 
and also implies the relation 
$$ 
\varrho_+(\omega) = \varrho_-(-\omega)\;,
$$
between the spectral functions $\varrho_\pm$ of the two dot levels. 
[For a precise definition, see Eq.~\eqref{eq:spectralfunctions}]

Finally, at the special point, $\epsilon_+=\epsilon_- =0$, we have yet
another electron-hole symmetry, satisfied even for 
$|t_{\alpha +}| \ne |t_{\alpha -}|$,
\bea 
(d^\dagger_{\alpha,\uparrow},\;d^\dagger_{\alpha,\downarrow}) &\leftrightarrow &
(d_{\alpha,\downarrow},\;-\;d_{\alpha,\uparrow})\;,
\nonumber
\\
(c^\dagger_{\xi, L \uparrow},\;c^\dagger_{\xi,L\downarrow}) &\leftrightarrow &
(-\;c_{-\xi, R\downarrow},\;c_{-\xi,R\uparrow})\;.
\label{eh2}
\eea
This second symmetry impleas the spectral function relations, 
$\varrho_+(\omega) = \varrho_+(-\omega)$ and 
$\varrho_-(\omega) = \varrho_-(-\omega)$.

\subsection{Out of equilibrium iterative perturbation theory}\label{iptsec}

To describe the transport properties of a quantum dot under 
out of equilibrium conditions, 
we shall apply the so-called Keldysh Green function technique.\cite{rammer}
However, before doing so, let us further reorganize our Hamiltonian. 
Clearly, interactions can substantially shift the effective values of the dot
energies. Therefore, to account for this trivial but possibly large renormalization
effect, we apply a counterterm procedure: We take
as a non-interacting part the following Hamiltonian
\begin{equation}
   \tilde{H}_{0}\equiv H_{\rm cond}+H_{\rm hyb}+   
\sum_{i,\sigma} \tilde{\varepsilon}_{i\sigma}\;d_{i\sigma}^{\dagger}d_{i\sigma}\;,  
\label{eq:tildeH_0}
\end{equation}
while the rest of the Hamiltonian is treated as a perturbation, 
\bea
\tilde H_{\rm int}&\equiv & : H_{\rm int}: + H_{\rm count}\;, 
\label{eq:tildeH_int}
\\
H_{\rm count} &\equiv &
\sum_{i,\sigma}(E_{i}-\tilde{\varepsilon}_{i\sigma})d_{i\sigma}^{\dagger}d_{i\sigma}\;.
\eea
In other words, we expand around  fictitious (effective) dot
levels, $\tilde{\varepsilon}_{i\sigma} $, at the price of  treating
 the counterterm  $H_{\rm count}$ also as a perturbation. This approach is
essentially the same as 
performing perturbation theory around the Hartree theory, and allows
us to  extend the range of validity of our perturbation theory 
by making a proper choice of $\tilde{\varepsilon}_{i\sigma}$ (see below).

Our primary goal is to compute the non-equilibrium Keldysh Green's functions,
$G^{\kappa,\kappa'}$, with the index pair $(\kappa,\kappa')$ 
corresponding to the branches of the Keldysh contour. With this notation
$G^{1,1}$ stands for time-ordered ($G^T$), $G^{1,2}$  for lesser ($G^<$), 
$G^{2,1}$ for larger ($G^>$), and $G^{2,2}$ for the anti-time-ordered 
($G^{\tilde T}$) Green's functions. 
These are related to the usual retarded ($R$), advanced ($A$) and 
Keldysh ($K$) Green's functions by the usual relations\cite{rammer}
\begin{eqnarray}
  G^{R} & = & G^{T}-G^{<}\;,\label{geret}\\
  G^{A} & = & G^{T}-G^{>}\;,\\
  G^{K} & = & G^{T}+G^{\tilde{T}}\;,\label{gkel}
\end{eqnarray}
and also satisfy the restriction, $G^{T}+G^{\tilde{T}}=G^{<}+G^{>}$.

To determine the interacting dot Green's functions, 
$G_{i\sigma,i'\sigma'}^{\kappa\kappa'}$, we perform perturbation theory 
in ${\tilde H}_{\rm int}$ by taking ${\tilde H}_0$ as a non-interacting 
Hamiltonian.  The Green's function can thus be expressed in terms of the 
self-energy $\Sigma_{i\sigma,i'\sigma'}^{\kappa\kappa'}$ and the unperturbed Green's functions, 
$g_{i\sigma,i'\sigma'}^{\kappa\kappa'}$ by the Dyson equation,
\begin{equation}
   \mathbf{G}(\omega)^{-1} = \mathbf{g}(\omega)^{-1}-\mathbf{\Sigma}(\omega)\label{dys}\;.
\end{equation}
where a matrix notation $A_{i\sigma,i'\sigma'}^{\kappa\kappa'}\to  \mathbf{A}$ 
has been introduced. The non-interacting Green's functions corresponding 
to the resonant level model ${\tilde H}_0$ can be easily 
computed, and are listed in Appendix~\ref{firstapp}.
Up to second order in the interaction, the self-energy
can formally be written as 
\begin{equation}
   \mathbf{\Sigma}(\omega)=
   \mathbf{\Sigma}^{\rm (count)}
  +\mathbf{\Sigma}^{(1)}
  + \mathbf{\Sigma}^{(2)}(\omega)+\dots
  \;,\label{selfen}
\end{equation}
where   the different terms denote  the counterterm, the first and second order terms shown in Fig.~\ref{self}.
An explicit evaluation of the diagrams in Fig.~\ref{self}
gives the following results: 
\begin{eqnarray}  
& &{\Sigma^{\rm (count)}}_{ii'\sigma}^{\kappa\kappa'}
= \delta_{\kappa\kappa'}\delta_{ii'}s_{\kappa}
  \left(E_{i}-\tilde{\varepsilon}_{i\sigma}\right)\;,\nonumber\\
  & &{\Sigma^{(1)}}_{ii'\sigma}^{\kappa\kappa'}
      =\delta_{\kappa\kappa'}s_{\kappa}\int\limits_{-\infty}^{\infty}
  \frac{d\omega_{1}}{2\pi}\sum_{jj'\sigma'}\Gamma_{i\sigma\;j\sigma'}^{i'\sigma \;j'\sigma'}g_{jj'\sigma'}^{<}(\omega_{1})\nonumber\\
  & &{\Sigma^{(2)}}_{\;ii'\sigma}^{\kappa\kappa'}(\omega) = \nonumber\\ 
   & = & s_{\kappa}s_{\kappa'}\sum_{\substack{j,m,n\\j',m',n'\\\sigma', \sigma'',\sigma'''}}\Gamma_{i\sigma\;n\sigma'''}^{j\sigma'\;m\sigma''}
   \Gamma_{j'\sigma'\;m\sigma''}^{i'\sigma \;n'\sigma'''}
  \int\limits_{-\infty}^{\infty}\frac{d\omega_{1}}{2\pi}\int\limits_{-\infty}^{\infty}\frac{d\omega_{2}}{2\pi}\times\nonumber\\
  & \times & g_{jj'\sigma'}^{\kappa\kappa'}(\omega_{1})\;g_{mm'\sigma''}^{\kappa\kappa'}(\omega_{2})\;
   g_{n'n\sigma'''}^{\kappa'\kappa}(\omega_{1}-\omega_{2}-\omega)
\label{eq:self_exp}
\;,
\end{eqnarray}
where the Keldysh sign $s_{\kappa}$ has been introduced to account 
for the sign change of the interaction term on the Keldysh contour: 
$s_{1}=1$, $s_{2}=-1$. In these expressions we assumed that 
the $z$-component of the spin is conserved, and correspondingly, 
the Green's functions and the self-energies are diagonal in $\sigma$.

In principle, Eqs.~ (\ref{dys}), (\ref{selfen}), and (\ref{eq:self_exp}) 
give a complete perturbative
description of the quantum dot. However, they depend parametrically on
the so far unspecified levels, $\tilde{\varepsilon}_{\pm\sigma}$. 
Following the concept of iterative perturbation theory, we determine these quantities 
selfconsistently from Eqs.~(\ref{selfen})
and (\ref{dys})
by requiring that ${\bf g}$ and ${\bf G}$ give the same occupation numbers,\cite{Yeyati}
\begin{eqnarray} 
n_{i\sigma}^{(0)}\left[ \tilde{\varepsilon}_{i\sigma}\right] 
& \equiv & n_{i\sigma}^{(2)}\left[ \tilde{\varepsilon}_{i\sigma}\right]
\label{condi},
\\
   n_{i\sigma}^{(0)}\left[ \tilde{\varepsilon}_{i\sigma}\right] & = & 
   \frac{1}{2\pi i}\int\limits_{-\infty}^{\infty}g_{ii\sigma}^{<}(\omega)\;d\omega\;,\nonumber\\ 
  n_{i\sigma}^{(2)}\left[ \tilde{\varepsilon}_{i\sigma}\right] & = & 
\frac{1}{2\pi i}\int\limits_{-\infty}^{\infty}G_{ii\sigma}^{<}(\omega)\;d\omega \nonumber\;.
\end{eqnarray}

Condition (\ref{condi}) can be fulfilled by
tuning the effective levels. In spite of the simplicity of 
this ``iterative perturbation theory'', 
the results obtained with it show remarkable agreement 
with the  ones obtained with 
more advanced techniques for several equilibrium systems.\cite{Schoeller}
Unfortunately, a well established and exact non-equilibrium 
impurity solver being not available yet, one can only 
judge the validity of this perturbative approach by possibly 
comparing with results obtained by some reliable equilibrium methods, 
and by investigating the internal consistency of its. As discussed in Section \ref{numerics}, the range 
of applicability turns out to be limited to small to moderate  values 
of $U$ and $J$. 

In principle, Eq.~\eqref{condi} could and 
should be solved for finite biases. However, 
under non-equilibrium conditions, the off-diagonal elements 
$\langle d^\dagger_{1\sigma} d_{2\sigma}\rangle$ become also finite, 
and additional constraints 
related to current conservation may also emerge.  In this work, 
we rather use
a different and simple strategy: we determine ${\tilde \varepsilon}_{i}$ in 
equilibrium, and then fix it also for the non-equilibrium calculations. 
For not very large bias voltages this leads to a stable 
solution of the self-consistency equations.

\subsection{Transport properties}\label{transport}

Having the Green's functions ${\bf G}$ at hand, 
we can calculate transport properties such as current and conductance 
based upon the Meir--Wingreen formula \cite{meir},
\begin{eqnarray}
   I & = &\frac{ie}{2h}\sum_{\sigma}\int\limits_{-\infty}^{\infty} d\omega\mathrm{Tr}\Big[(\mathbf{\Gamma}^{L}
-\mathbf{\Gamma}^{R})\mathbf{G}_{\sigma}^{<}(\omega)+\nonumber\\
& + &  
(f_{L}(\omega)\mathbf{\Gamma}^{L}-f_{R}(\omega)\mathbf{\Gamma}^{R}) 
(\mathbf{G}_{\sigma}^{R}(\omega)-\mathbf{G}_{\sigma}^{A}(\omega)) \Big]\;.\label{curexp}
\end{eqnarray}
Here $f_{\alpha}(\omega)=f(\omega-\mu_{\alpha})$ is the Fermi function in lead $\alpha$,  and the Green's functions are matrices in the level indices only. 
The difference between the formulas for lateral and vertical dots 
appears only in the different structure of the $\mathbf{\Gamma}$-matrices 
defined as
\begin{equation}
   (\Gamma_{ij}^{\alpha})_{\rm lat} =2\pi\; N_{0}\;t_{\alpha i}t_{\alpha j}^{*}\;, 
\end{equation}
in the lateral case and
\begin{equation}
   (\Gamma_{ij}^{\alpha})_{\rm vert} = \delta_{ij} \;2\pi \; N_{0}\;t_{\alpha i}t_{\alpha j}^{*}\;, 
\end{equation}
for a vertical quantum dot. These are related to the width
$\Gamma_\pm$ of levels $i=\pm$ through 
$$
\Gamma_i \equiv \sum_\alpha \Gamma_{ii}^{\alpha}\;.
$$
 
In our case, the ground state of the system is a Fermi liquid
\cite{hewson}. 
Therefore, according to Nozi\'eres's Fermi liquid theory, 
at $T=0$ temperature (and 
$V=0$ bias),  quasi-particles at the Fermi energy scatter elastically 
from the impurity, and their scattering process can be described in terms 
of simple phase shifts. Since our model is invariant under reflection 
for $V=0$, and since one of the levels is assumed to be even while the other 
one odd, we can characterize the scattering process at the Fermi energy 
by just two phase shifts, $\delta_i$, associated with the two dot levels. 
These phase shifts are related to the retarded equilibrium Green's 
function  through the Fermi liquid relation,\cite{hewson}
\begin{equation}
 \delta_{i}=\frac{\pi}{2}
-\arctan\left(\frac{\mathrm{Im}G_{ii\sigma}^{R}(\omega=0,V=0)}
{\mathrm{Re}G_{ii\sigma}^{R}(\omega=0,V=0)}\right)\;.
\label{delta_form_G}
\end{equation}
In the limit of infinite bandwidth, the phase shifts 
 are also related to the occupation of the levels $d_{i\sigma}$ 
by the Friedel sum rule,\cite{hewson} 
\be
2\frac {\delta_i}\pi = \langle n_i\rangle \;.
\label{delta_form_n}
\ee 
where $\langle n_{i}\rangle =\sum_{\sigma}\langle n_{i\sigma}\rangle $.
However, for a finite conduction band cut-off Eq.~\eqref{delta_form_n} is only 
approximate, and  Eq.~\eqref{delta_form_G} gives a more reliable way 
to determine the phase shift.

The $T=0$ linear conductance is directly related to the phase 
shifts above through the Landauer-B{\" u}ttiker formula. 
For a lateral dot, an elementary calculation gives\cite{Pustilnik}
\begin{equation}
  G_{\rm lin}^{\rm lat} =
  \frac{2e^2}{h}\sin^{2}(\delta_{+}-\delta_{-})\label{glinlat}\;, 
\end{equation}
while for a vertical dot one obtains\cite{Pust2}
\begin{equation}
   G_{\rm lin}^{\rm ver}=
\frac{2e^{2}}{h} \left(\sin^{2}\delta_+
+ \sin^{2}\delta_- \right)\;. 
\label{glinver}
\end{equation}
These equations are very instructive, and help us to understand the transport 
properties of the dots. They imply, {\em e.g.}, that at complete resonance, 
$\delta_+=\delta_-=\pi/2$, the conductance of a vertical dot is $4e^2/h$, 
while that of a lateral dot vanishes due to the interference of 
scattering states. These interference effects are expected to show 
up also in the non-equilibrium results.

The equations above are also useful to test our  numerical calculations. 
The linear conductance values in 
Eq.~(\ref{glinlat}) and in  Eq.~(\ref{glinver}) can be compared to the 
equilibrium limits of the differential conductances 
obtained by a  numerical differentiation of the current as a function of 
the bias voltage $\mathrm{d}I/\mathrm{d}V|_{V=0} $, as computed 
from the Meir-Wingreen formula.

\subsection{Computational details}\label{compsec}

In the numerical calculations we represented the 
 Green's functions using  a 
finite uniform mesh  of $N$ frequency  points 
in the range $-\Omega/2<\omega<\Omega/2$. 
Depending on the accuracy needed, we used slightly different values 
for $\Omega$ and $N$. To study the range of applicability of our method, 
we used  $N=2^{15}$ points and $\Omega=500U$.
To compute the spectral functions, we used $N=2^{17}$ and 
$\Omega=1000U$, while for the current and  differential 
conductance plots we chose $N=2^{17}$ and $\Omega=200U$. 
All results presented in this paper were obtained 
at $T=0$ temperature. 

The advantage of the iterative perturbation theory is 
that the calculation of the second order self-energy 
is only  seemingly cumbersome. Rewriting   
Eq.~(\ref{eq:self_exp}) 
in time domain, one obtains a simple multiplication of the 
various components of the Green's functions, 
\begin{eqnarray}
   {\Sigma^{(2)}}_{\;ii'\sigma}^{\kappa\kappa'}(t) & = & 
   s_{\kappa}s_{\kappa'}\sum_{
\substack{j,j',m,m',n,n'\\\sigma', \sigma'',\sigma'''}}
\Gamma_{i\sigma\;n\sigma'''}^{j\sigma'\;m\sigma''}
   \Gamma_{j'\sigma'\;m\sigma''}^{i'\sigma \;n'\sigma'''}\times\nonumber\\
   & \times &\;g_{jj'\sigma'}^{\kappa\kappa'}(t)\;g_{mm'\sigma''}^{\kappa\kappa'}(t)\;
   g_{n'n\sigma'''}^{\kappa'\kappa}(-t)\label{2ndself2}\;. 
\end{eqnarray}
Thus the self-energy can be computed by just performing  
a  Fast Fourier Transformation to get ${\bf g}(t)$, evaluating 
${\bf \Sigma}(t)$ in the time domain, and then transforming it back to
frequency space. 

As already mentioned before, for any given set of parameters 
we first carried out a calculation in the absence of voltage, 
$V=0$, and determined the levels, $\tilde \varepsilon_\pm$. 
Then we computed ${\bf \Sigma}(\omega)$ and ${\bf G}(t)$
using Eq.~\eqref{2ndself2}, 
computed the spectral functions 
\begin{equation}
   \varrho_{i\sigma}(\omega)=-\frac{1}{\pi}\mathrm{Im}\;G_{\;ii\sigma}^{R}(\omega)\;,
\label{eq:spectralfunctions}
\end{equation}
as well as the  total spectral function, 
 $\varrho_{T}(\omega)=\sum_{i\sigma}\varrho_{i\sigma}(\omega)$, 
and finally  evaluated the current 
using the Meir-Wingreen formula, Eq.~\eqref{curexp}. 
The differential conductance has been obtained by direct numerical
differentiation of the current, $G(V) = dI(V)/dV$.

\subsection{Limitations}\label{numerics}

\begin{figure}
\includegraphics[width=240pt,clip=true]{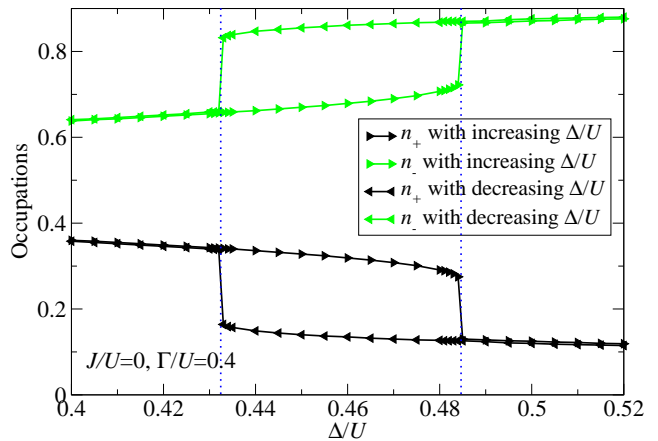}
\caption{(Color online) Hysteresis of the occupation number as a
  function of level splitting ($\Delta$) for  $J/U=0.0$  
and $\Gamma_{\pm}/U=\Gamma/U=0.4$.}\label{hist}
\end{figure}
Perturbation theory has a limited range of validity. 
For large values of the interaction, $U$, a spontaneous 
symmetry breaking occurs, whereby the occupation of the dot levels 
$i=\pm$ becomes unequal  even for $\Delta=\epsilon_+-\epsilon_-=0$, 
$\langle n_+\rangle \ne \langle n_-\rangle $. 
This phenomenon is similar to the spontaneous moment formation 
found by Anderson,\cite{Anderson,BerciPRB} which we eliminated here by
setting the Green's functions explicitly proportional 
to $\sim \delta_{\sigma\sigma'}$.
In a perturbative calculation
one should always stay away from these regions in  parameter 
space to avoid spurious  solutions. Fortunately, as shown in Fig.~\ref{hist}, we
can easily identify these
"dangerous" regions  by  computing   the occupation 
numbers $\langle n_\pm\rangle$
as a function of level splitting, $\Delta/U$, since   
$\langle n_\pm\rangle$ exhibit hysteresis there. 

\begin{figure}
\includegraphics[width=7cm,clip=true]{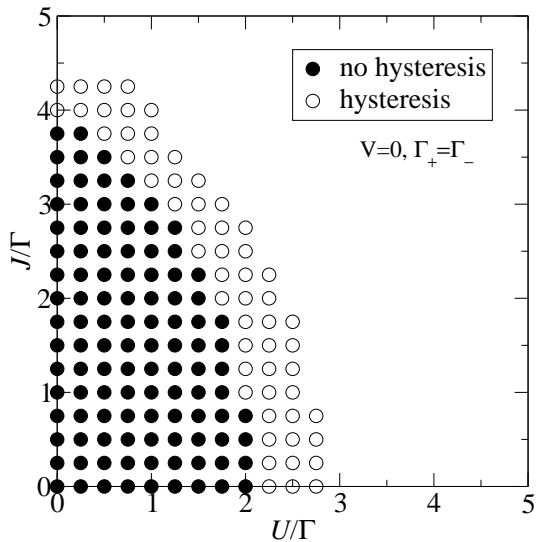}
\caption{Stability region of perturbation theory  
 for $\Gamma=\Gamma_{\pm}$.  Empty circles mark the region of 
hysteresis, while  filled circles indicate the stable region, where PT gives 
meaningful results.  
}\label{phdiag1}
\end{figure}

Carrying out a similar hysteresis analysis for each value of 
the parameters 
$U/\Gamma$ and  $J/\Gamma$, we  obtain 
the stability diagram, Fig.~\ref{phdiag1},  
delineating the region of applicability of our perturbative 
approach (for simplicity, 
here we took  $\Gamma_\pm =\Gamma$). Clearly, one needs  to keep both
$U/\Gamma$ and $J/\Gamma$ moderate to stay within the range of
applicability of  iterative perturbation theory. 

\section{Results and Discussion}

\subsection{The symmetric case $(\Gamma_{+}=\Gamma_{-})$}\label{sym}

The parameter space of the two-level quantum dot is huge. Therefore, 
to get more intuition, we first restrict somewhat the number of
independent parameters by assuming that
both levels couple to the leads with equal strength, 
$\Gamma_\pm \equiv \Gamma$. We assume further that the dot is at the 
particle-hole symmetrical point, $\epsilon_+=-\epsilon_- = \Delta/2$, 
where the number of electrons on the dot is fixed to 
$n= \sum_{i,\sigma}\langle n_{i\sigma}\rangle =2 $.
Before presenting the numerical data, let us briefly discuss 
the structure of the ground state as a function of $J$, $U$, and
$\Delta$ in case of small  tunnelings, $\Gamma$.

\begin{figure}
\begin{center}
\includegraphics[width=170pt,clip=true]{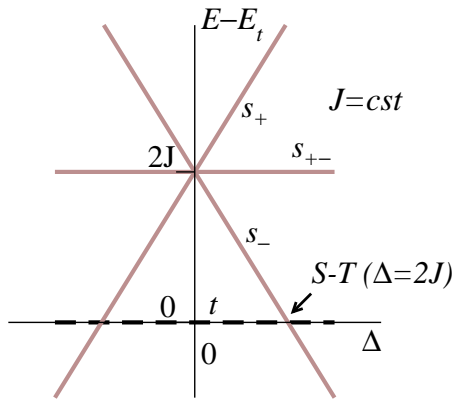}
\end{center}
\caption{
Evolution of the singlet and triplet states of the isolated dot
for fixed $J>0$, as a function of $\Delta$. Energies are measured
from the energy $E_t$ of the triplet state. 
The singlet-triplet  transition point is also indicated (S-T). 
}\label{states}
\end{figure}

In the absence of tunneling to the leads, $\Gamma=0$, the states of
the  dot can be grouped into three singlets (with spin $S=0$)
and a triplet $(S=1)$. Two singlet states, $|s_{\pm}\rangle$ 
correspond to putting two electrons on the upper/lower
level, while a third state, $|s_{+-}\rangle$, is a singlet  with 
one electron residing on each level. The rest of the two-electron
states are spanned by a triplet state, $|t,{m}\rangle$ ($m=0,\pm$), 
where the spin of the electron on level $i=+$ is aligned 
with that on level $i=-$. Fig.~\ref{states} shows the evolution of these four levels as a function 
of the splitting $\Delta$ for finite 
$J$. For $\Delta=0$ the triplet state is the ground state, and the 
three singlet states form degenerate  excited states of energy $2J$. 
Increasing $\Delta$, however, pulls down the state $|s_{-}\rangle$, 
which becomes the ground state for $\Delta > 2J$.

\begin{figure}[b]
\begin{center}
\includegraphics[width=220pt,clip=true]{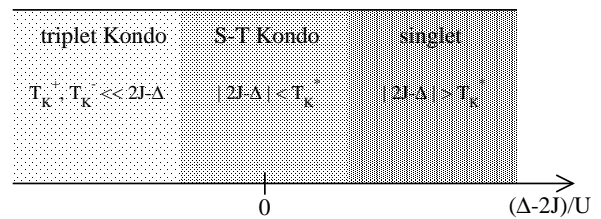}
\end{center}
\caption{Sketch of the singlet, triplet, and transition 
 regimes, as a function of $\Delta/U$ for a fixed $J>0$. 
}
\label{states_delta}
\end{figure}

This picture is slightly modified if the dot is coupled to the
leads.  In this case, for generic couplings,\cite{Pustilnik,hofstetter} 
the transition becomes a smooth cross-over, so that the triplet and
singlet states  are adiabatically connected.  However, the various
regimes are described by rather different physical pictures. 
For large values of $\Delta$, the two dot electrons form a local
singlet on the dot, while for  $J\ll |\Delta|$ they are
aligned into a triplet, which then couples to the leads through an
exchange coupling, and is screened by a two-stage Kondo effect
for $\Gamma_+<\Gamma_-$ at some temperatures $T_K^+ < T_K^-$.
In this regard, the case $\Gamma_-\equiv\Gamma_+$ is rather special, 
since then  only a single Kondo scale appears, $T_K^+=T_K^-$.
As one approaches the transition region, $\Delta \approx 2 J$,  
from the triplet side, the Kondo temperatures $T_K^\pm$
increase, until  the singlet state $s_-$ and the  triplet states are all
inseparably mixed into a single quantum state, characterized 
by  a Fermi liquid scale, $T_K^*$.\cite{eto,avishai} 
These three  regions are sketched in  Fig.~\ref{states_delta}.

\subsubsection{Spectral functions}\label{specsec}

To gain insight to the effects governing the 
conductance through the dot, it is useful to study  
first the equilibrium spectral functions.

\begin{figure}
\includegraphics[width=240pt,clip=true]{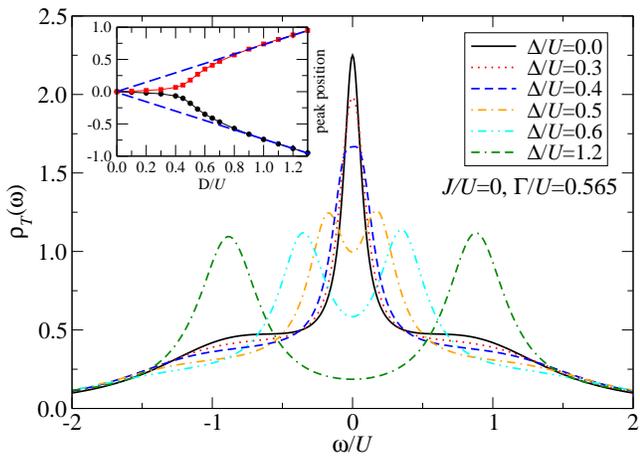}
\caption{Total equilibrium spectral functions, $\varrho_{T}(\omega)$,  for  $J/U=0$ and 
$\Gamma_{\pm}/U=\Gamma/U=0.565$, and different values of the 
level splitting, $\Delta/U$. 
Inset: Evolution of the peak positions in $\varrho_{\pm}(\omega)$
 as a function of $\Delta/U$. The  
dashed lines indicate $\omega=\pm \Delta$ 
}
\label{speceq}
\end{figure}

\emph{Case of no Hund's rule coupling, $J\equiv0$.}
Let us first investigate in the absence  of Hund's rule coupling,
$J\equiv 0$. 
In this case, there is no triplet region, 
rather, changing   $\Delta$, one finds a transition between 
the two singlet states 
 $|s_-\rangle\to |s_+\rangle$ through a peculiar Kondo state 
formed by all three singlet states and the triplet (see Fig.~\ref{states}).
Fig.~\ref{speceq} shows the total  spectral function,
$\varrho_T(\omega)$, in this case, for  $\Gamma/U=0.565$, as 
a function   of level splitting, $\Delta/U$.

For $\Delta/U=0$ and $J/U=0$, the ground state of an isolated dot is sixfold degenerate. 
When the quantum dot is connected to the leads, a  
strong  Kondo resonance driven by the quantum fluctuations of this 
sixfold degenerate state appears at $\omega = 0$. The width of this 
peak can be identified as the Kondo temperature, $T_K^*$. 
In addition, one
observes  two shoulders at $\omega\approx\pm U$, which can be
identified as the Hubbard peaks. 

Increasing $\Delta$ first slightly suppresses the Kondo 
resonance at $\omega=0$, and finally, for $\Delta> T_K^*$ splits 
the resonance into two side-peaks. These side-peaks can be identified
as the singlet-triplet excitation  peaks, 
 and their position  can be precisely determined from the 
level-projected spectral
functions, $\varrho_\pm(\omega)$. As shown in the inset of 
Fig.~\ref{speceq}, for large values of $\Delta$ they are indeed
located at $\omega\approx \pm \Delta $, although the splitting 
takes place rather abruptly at $\Delta\approx 0.4$, and the 
$\Delta-$dependence of 
the maxima of $\varrho_\pm(\omega)$  is very non-linear.

\begin{figure}[b]
\begin{center}
\includegraphics[width=240pt,clip=true]{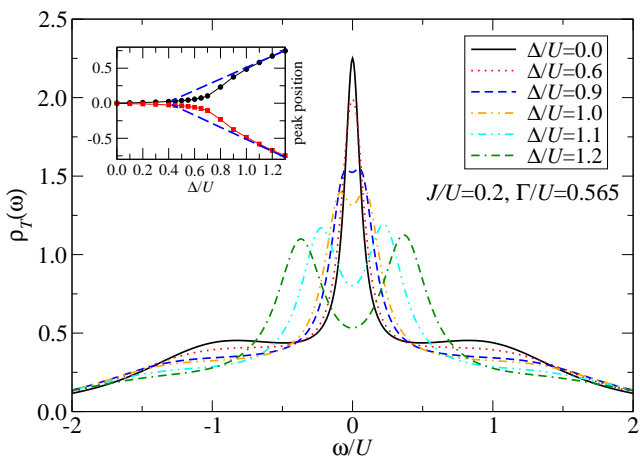}
\end{center}
\caption{
Total equilibrium spectral functions, $\varrho_{T}(\omega)$, 
for $J/U=0.2$, $\Gamma_{\pm}/U=\Gamma/U=0.565$, for
 different values of level splitting, $\Delta/U$. 
Inset: Peak positions of  $\varrho_{\pm}(\omega)$
 as a function of $\Delta/U$. The dashed lines indicate 
$\omega=\pm (\Delta-2J)$.} 
\label{speceqj=02}
\end{figure}

\emph{Finite Hund's rule coupling, $J>0$.}
The effect of the level splitting $\Delta$ on the equilibrium spectral 
function  is presented in   Fig.~\ref{speceqj=02}  for  parameters $J/U=0.2$ 
and $\Gamma/U=0.565$. 
The spectral functions behave quite similarly to those calculated for 
$J=0$: A Kondo resonance appears for $\Delta=0$ in 
$\varrho_T(\omega)$, which is gradually suppressed as $\Delta$ increases, and finally 
splits into two resonances for large values of $\Delta$. 
However, in this case, we should interpret the Kondo resonance slightly
differently than before. For $\Delta=0$ the ground state of the
isolated dot would be a triplet state, and therefore turning on 
a small $\Gamma$, this spin
triplet gets screened by the conduction electrons in the leads through
a {\em triplet Kondo effect}.\cite{izumida2} Although $\Gamma$ is relatively
large in our case and  comparable to the singlet triplet splitting, 
$\Gamma\sim 2J$, nevertheless, this interpretation still holds, since
the width of the central resonance ($T_K$) is still smaller 
 than the splitting $\Delta E_{st}\equiv |\Delta - 2J|$.
 In fact, the resonance splits 
approximately where $\Delta E_{st}  \sim T_K$, and
the positions of the split resonances follow approximately the
straight lines $\omega \approx \pm (\Delta - 2 J)$ for  large $\Delta$'s (see
the inset of Fig.~\ref{speceqj=02}), in complete agreement with the ST
transition picture.

\subsubsection{Conductance}\label{consec}

\begin{figure}[t]
\includegraphics[width=240pt,clip=true]{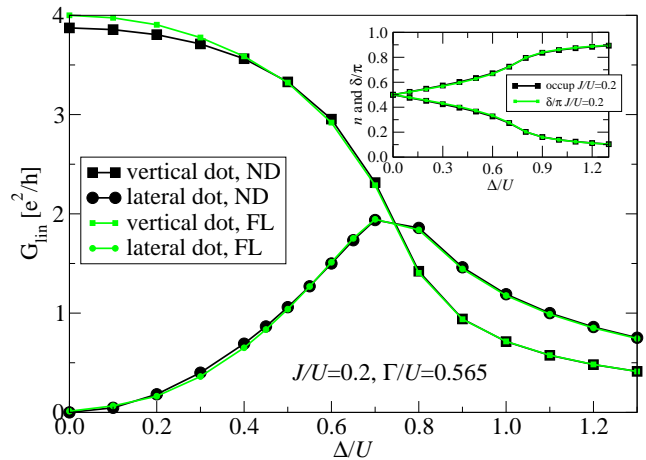}
\caption{
Linear conductance of lateral and vertical dots 
for $J/U=0.2$, from numerical derivation (ND) of the current at zero bias, 
 and from the Fermi liquid relations (FL). Inset:  Comparison of 
occupation  numbers and phase shifts, $\delta/\pi$. The conductance of
 a symmetrical  lateral dot shows a maximum of height $2e^2/h$ around
 the ST transition.} 
\label{glin1}
\end{figure}

Having investigated the equilibrium spectral functions, let us now
turn to the discussion of non-equilibrium transport through the dot. 
We first focus on the {\em linear conductance}, 
$G_{\rm lin}={\rm d}I/{\rm d}V|_{V=0}$. 
This is plotted in  Fig.~\ref{glin1}  as a function of $\Delta$
both for lateral and vertical
dots with a finite Hund's rule coupling,
$J/U=0.2$. We 
computed $G_{\rm lin}$, on one hand,  by numerically differentiating 
the current, Eq.~(\ref{curexp}), but we also computed it 
by extracting  the phase shifts from the retarded Green's functions
and the  occupation numbers, and then using the  Landauer-B{\" u}ttiker formula 
(Eq.~(\ref{glinlat}) for lateral and Eq.~(\ref{glinver}) 
for vertical dots). 
As shown in Fig.~\ref{glin1}, 
the two procedures give identical results within
numerical accuracy.

Vertical and lateral dots exhibit very different
characteristics. The linear conductance of a lateral dot shows 
a maximum around the ST transition and it is small on both sides of
the transition, while the conductance of a vertical dot 
crosses  over smoothly from a conductance of $G_{\rm lin}\approx 4e^2/h$ at 
$\Delta=0$ to a small value for $\Delta\gg 2J$. This behavior can be
understood from  Eqs.~(\ref{glinlat}) and (\ref{glinver}), and
the behavior of the phase shifts, shown in the inset of 
 Fig.~\ref{glin1}: In the triplet regime,
 $\Delta\approx0$, two conduction electrons 
are needed to screen the local spin $S=1$. Therefore, both phase
shifts are close to $\delta_\pm\approx \pi/2$ by the Friedel sum
rule.\cite{Hewson} While for a vertical dot the contributions of the 
 channels  $i=\pm$ add up to the conductance, and amount in a total
 conductance of $G_{\rm lin}=4e^2/h$, for lateral dots there
 is a destructive interference (see Eq.~\ref{glinlat}), which finally
 amounts in a complete back-reflection of electrons and no conductance, 
$G\approx 0$.\cite{pusgla,hofstetter,Zarand}
  Increasing $\Delta$, the phase shifts gradually cross
 over to values, $\delta_- \approx \pi$ and $\delta_+ \approx 0$. This gives rise to a
 maximal conductance of $G_{\rm lin}=2e^2/h$ somewhere
in the vicinity of the ST transition for a lateral dot,\cite{pusgla}
while it results in the monotonous decrease of the conductance 
of a vertical dot,\cite{izumida2} as shown in Fig.~\ref{glin1}.
The Friedel sum rule, Eq.~\eqref{delta_form_n}, is verified
in the inset of Fig.~\ref{glin1}, where we compare the phase shifts 
to the occupation numbers.

\begin{figure}
\includegraphics[width=240pt,clip=true]{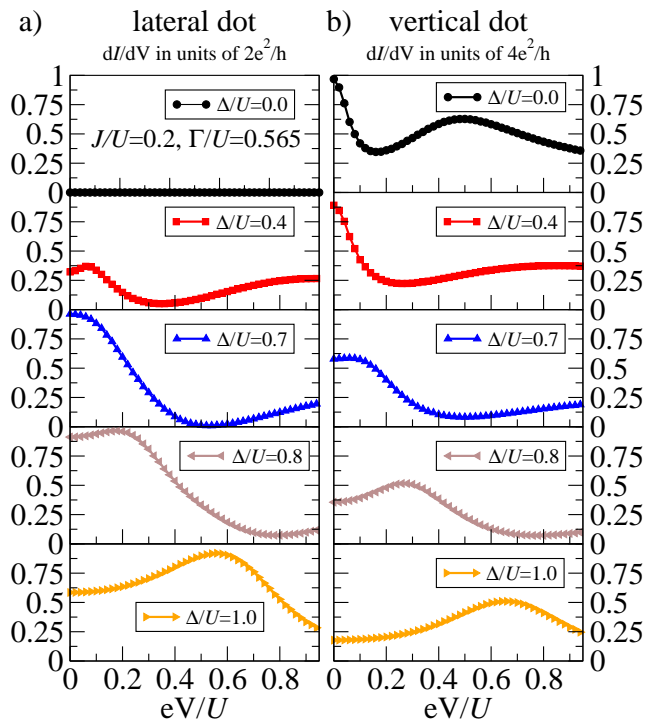}
\caption{Differential conductance, $G(V)$, for $J/U=0.2$ and 
$\Gamma_{\pm}/U=\Gamma/U=0.565$ for different level splittings, $\Delta/U$ 
both for  lateral (a) and for  vertical (b) quantum dots. For vertical dots 
 the suppression and the splitting  of the Kondo resonance 
are clearly visible in  $G(V)$. For lateral dots, 
interference effects mask the signal on the triplet side.
}
\label{didveqj=02}
\end{figure}
The differential conductance, $G(V)$, can be observed in Fig.~\ref{didveqj=02}.
For a vertical dot, the behavior of $G(V)$ follows 
the naive expectations  based upon the spectral functions' structure 
and the excitation spectra of the dots: For $\Delta=0$ a clear Kondo resonance
is seen at $V\approx 0$, and the singlet excitations give a broad
resonance  at $V\approx 2J/e$. Increasing $\Delta$, the central resonance is
suppressed in amplitude (as predicted by the Friedel sum rule), but it also
broadens
somewhat as one approaches the ST transition point, where quantum
fluctuations give rise to a somewhat higher Kondo temperature,
$T_K^*$.\cite{eto,avishai} The excitation peak at $V\approx (2J-\Delta)/e$
shifts to lower energies and gradually merges with the central resonance.
Finally, for $\Delta> 2J$ the quantum dot is in the singlet
region. There the central resonance splits up and a
resonance is observed at $V\approx (\Delta-2J)/e$.

While it is easy to understand 
the differential conductance of a vertical dot, 
the conductance of a lateral dot is counter-intuitive. 
For $\Delta=0$ there is a complete destructive 
interference, and $G(V)\equiv 0$. We emphasize that 
this is only valid for the symmetrical case, $\Gamma_+ = \Gamma_-$, 
studied in this section. Increasing $\Delta$, $G(V)$ becomes non-zero,
and the conductance gradually increases. The shape of $G(V)$, however, does not have 
a simple intuitive explanation on the triplet side of the transition, 
as a result of the cancellation of the various contributions. 
Around the transition
point and beyond that ($\Delta> 2J$), however, the phase shifts are far 
away from $\pi/2$, interference effects are suppressed, 
and the conductance shows features which are quite similar to those
found for  vertical dots, discussed above.

\subsection{The asymmetric case, $\Gamma_{+}\neq\Gamma_{-}$}\label{assymsec}

\begin{figure}
\includegraphics[width=7cm,clip=true]{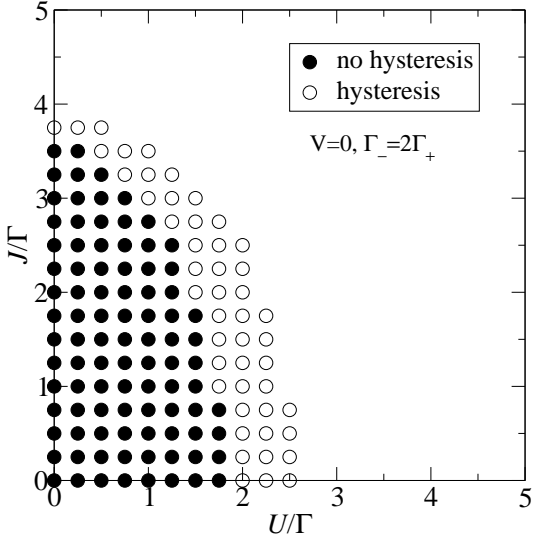}
\caption{
Range of applicability of perturbation theory 
for $\Gamma_{+}\neq\Gamma_{-}$, as extracted from the   
 hysteresis of occupation numbers. Filled circles indicate the region
 of stable perturbation theory. Couplings are measured in units 
of $\Gamma\equiv \sqrt{\Gamma_{+}\Gamma_{-}}$.} 
\label{phdiag2}
\end{figure}

In the previous subsection, we analyzed the case of equal tunneling
rates,  $\Gamma_{+}=\Gamma_{-}$. While this assumption  may 
be a good approximation for  some systems
(for carbon nanotubes, e.g.),\cite{hp1} 
it is violated for most quantum dots.\cite{Wiel,Granger} 
 Let us therefore investigate in this subsection 
the generic case of unequal couplings, $\Gamma_+\ne \Gamma_-$. 
The range of applicability  of our perturbative  method is 
shown in Fig.~\ref{phdiag2}. 
For simplicity, we fixed  $\Gamma_+ = \Gamma_-/2$  throughout this
subsection, and characterized the hybridization strength 
 in terms of the geometric mean of $\Gamma_\pm$,
$\Gamma\equiv\sqrt{\Gamma_{+}\Gamma_{-}}$. 
A comparison of   Fig.~\ref{phdiag2} and  Fig.~\ref{phdiag1}
 shows  that 
the range of applicability of perturbation theory  is only 
weakly modified by the assymetry in the tunneling rates.

For $\Gamma_+\ne \Gamma_-$, a few important differences appear though
compared to the case of equal tunneling strengths. 
When $\Gamma_{+}\neq\Gamma_{-}$,  
the particle-hole symmetry, Eq.~\eqref{eh1}, of the Hamiltonian 
is not valid anymore. As a consequence, 
the number of electrons on the dot is not exactly two 
and the spectral functions also violate the 
electron-hole symmetry relation, and $\varrho_+(\omega)\ne \varrho_-(-\omega) $.
Electron-hole symmetry is only maintained for 
$\Delta=0$, where 
 $\varrho_+(\omega)=\varrho_+(-\omega)$ and 
 $\varrho_-(\omega)=\varrho_-(-\omega)$ follow from the second electron-hole
symmetry transformation, Eq.~\eqref{eh2}. 
The second major difference is that, 
due to the two different coupling strengths, 
two separate Kondo scales emerge 
for the even and odd channels on the triplet side, 
$T_{K}^{\pm}$.\cite{Wiel}

\begin{figure}
\includegraphics[width=240pt,clip=true]{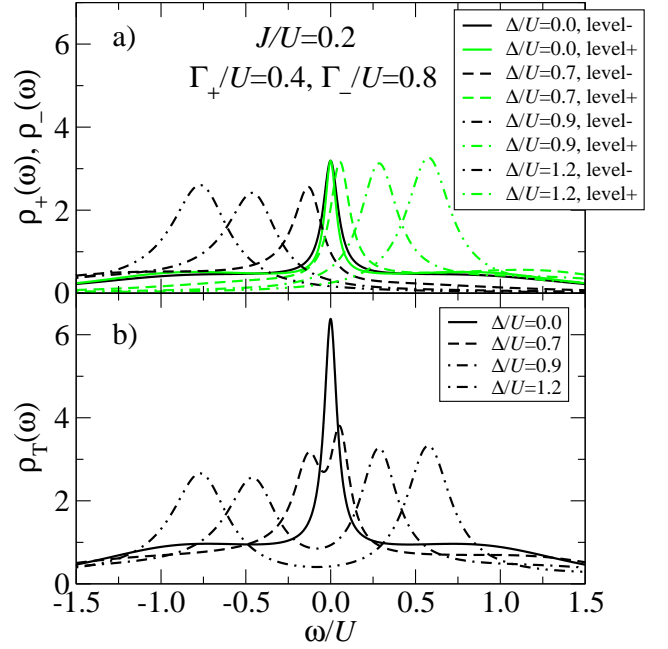}
\caption{Spectral functions for $J/U=0.2$, $\Gamma_{+}/U=0.4$ and $\Gamma_{-}/U=0.8$, for different values of level splitting, $\Delta/U$. 
The upper panel shows the level-projected 
 spectral functions, $\varrho_\pm(\omega)$, while in the 
lower panel the full spectral functions, $\varrho_T(\omega)$, are
presented. Notice the breaking of electron-hole symmetry for 
any $\Delta\ne 0$.}
\label{specnoneqGamma}
\end{figure}
Both differences are clearly visible in the spectral functions, 
shown in Fig.~\ref{specnoneqGamma}, where the level-projected 
spectral functions are also displayed. Clearly, only the 
$\Delta=0$ spectral functions are electron-hole symmetrical, 
and even there, the spectral functions $\varrho_+$ and $\varrho_-$
have a central Kondo resonance of different width. 

\begin{figure}[t]
\includegraphics[width=240pt,clip=true]{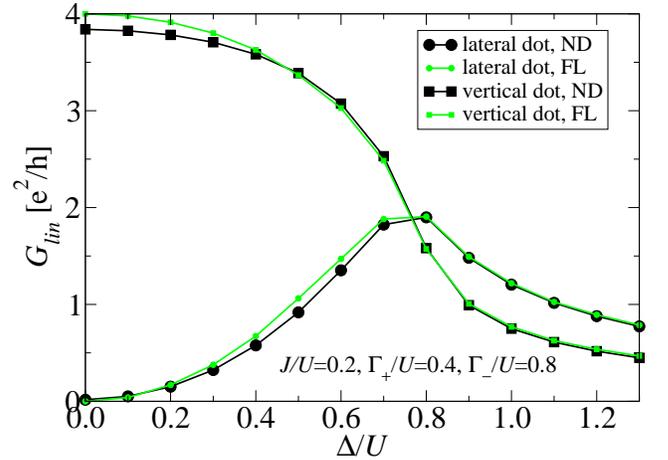}
\caption{Linear conductance for $J/U=0.2$, $\Gamma_{+}/U=0.4$ and
  $\Gamma_{-}/U=0.8$, as
 obtained from numerical derivation of the current (ND), and 
from the Fermi liquid relations (FL). The inset shows the phase shifts
$\delta_\pm/\pi$
extracted from the Green's functions at $\omega=0$,  and the occupation 
numbers, $n_\pm$.}\label{glin2}
\end{figure}

Apart from these differences, the overall evolution of the spectral
functions as well as  that of the phase shifts 
(see inset of Fig.~\ref{glin2}) is quite similar to the one observed for 
$\Gamma_+=\Gamma_-$. As a result, the behavior 
of the linear conductance
 as a function of the level splitting, $\Delta$, is also 
quite similar to the one obtained for $\Gamma_+=\Gamma_-$
(see Fig.~\ref{glin2}).

\begin{figure}[b]
\includegraphics[width=240pt,clip=true]{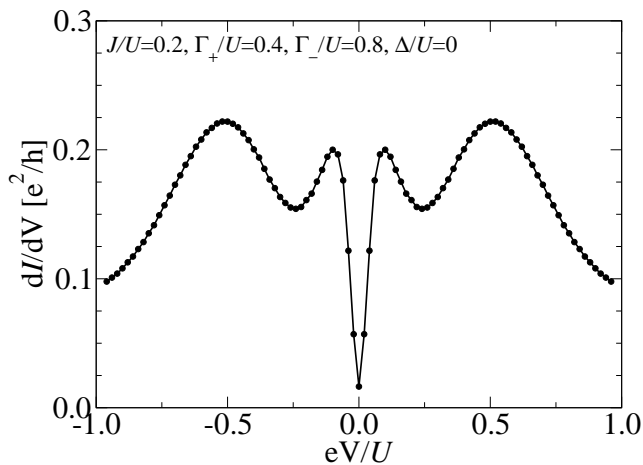}
\caption{Differential conductance 
of a  lateral quantum dot
for  $J/U=0.2$, $\Gamma_{+}/U=0.4$,
 and $\Gamma_{-}/U=0.8$, for level splitting $\Delta/U=0$.
The side-peak can be attributed to the singlet excitation energy, 
$2J$, while the central resonance-antiresonance 
structure is associated with the 
two consecutive Kondo effects on the triplet side.
}
\label{didvd=0}
\end{figure}

The \emph{differential conductance} is, however, quite different, and shows 
a much richer structure on the triplet side
in this case. For $\Delta=0$, both phase shifts are
pinned to $\delta_\pm=\pi/2$ by electron-hole symmetry. 
The physical reason for this is, of course the formation of a Fermi
liquid state through a Kondo effect in both channels on the triplet
side. 
As a consequence, by
Eq.~\eqref{glinlat}, 
the linear conductance vanishes also in this case, $G(V=0)=0$.
As we emphasized earlier, this result is due to 
an interference between the 
 channels, $i=\pm$. However, for 
 a voltage in the range, $T_K^+<eV<T_K^-$, the Kondo effect
in channel $i=+$ is destroyed, while 
the Kondo effect is still well-developed in channel  $i=-$. 
As a result, the destructive 
interference is suppressed as one turns on the voltage, and the conductance 
becomes non-zero. For even larger voltages,  
$T_K^-< eV$, the Kondo effect is also
destroyed in channel $i=-$, and the differential conductance starts 
to decrease
with increasing voltage. Thus the result of the consecutive
destruction of the two Kondo effects is the appearance of a  
 peak  at $eV\approx T_K^+$.\cite{Pustilnik,hofstetter,Zarand}
 This signature of the two-stage Kondo effect is clearly visible 
in the differential conductance, $G(V)$, shown in Fig.~\ref{didvd=0}.
There not only the central resonance-antiresonance
 structure can be seen, but also, 
a side-resonance at $\omega\approx 2J$, which can be associated with
the singlet excitations.

\begin{figure}
\includegraphics[width=240pt,clip=true]{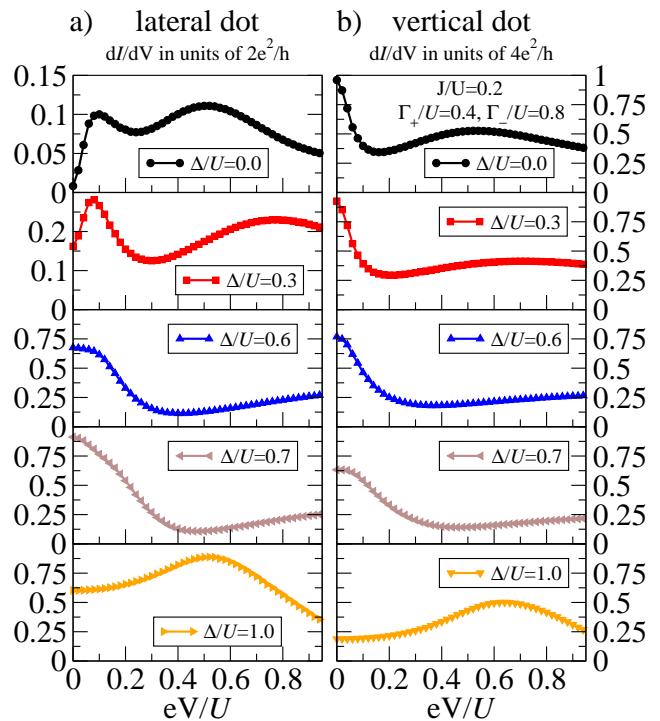}
\caption{Differential conductance, $G(V)$,  for $J/U=0.2$, $\Gamma_{+}/U=0.4$ 
and $\Gamma_{-}/U=0.8$ for different level splittings, $\Delta/U$ 
for a lateral (a) and a vertical (b) quantum dot.
The two consecutive Kondo resonances on the triplet side 
are clearly visible in the lateral dot (middle resonance-antiresonance
 structure), 
while the two Kondo resonances merge into a single peak in a vertical dot.}
\label{didvnoneqGamma}
\end{figure}

As shown in Fig.~\ref{didvnoneqGamma}, for the lateral dot, 
increasing $\Delta$  
gradually fills up the central dip, 
until the two resonances merge into a single ST-Kondo peak
at around $\Delta\approx 2J$, and finally split to two peaks 
located at $eV\approx \pm(\Delta-2J)$. All these features are in excellent
agreement with the experimental results of 
van der  Wiel \emph{et al.},\cite{Wiel}
and  Granger {\it et al.,} (see Fig.~3e in Ref.~\onlinecite{Granger}).

In Fig.~\ref{didvnoneqGamma} we also present our results 
for the differential conductance through a vertical dot. 
Contrary to the differential conductance of the lateral dot, 
these $G(V)$ curves do not exhibit any  remarkable difference 
with respect to the curves in Fig.~\ref{didveqj=02}, obtained 
for $\Gamma_+=\Gamma_-$. The main reason for this is that 
the perturbative approach does not estimate correctly the Kondo 
temperature: while for a non-perturbative calculation 
one would expect $T_K^+\ll T_K^-$ for the couplings used, perturbation 
theory  gives Kondo temperatures, $T_K^\pm$,
 of the same order of magnitude. 
While this is sufficient to produce a visible effect 
in the differential conductance of a lateral dot, it is 
not sufficient to see the expected  
 two-peak structure 
in the differential conductance of a vertical dot. 
In fact, for  $T_K^+\ll T_K^-$, the conductance should 
display two resonances on the top of each-other, a narrow one 
of width $\Delta V \sim T_K^+/e$ and a broad one 
of width $\Delta V \sim T_K^-/e$, 
both of  height $\Delta G\approx 2e^2/h$. However, since 
 $T_K^+\sim T_K^-$ in our calculation, these two peaks merge into a
single resonance of height $4e^2/h$ on the triplet side of the
transition.

Our theoretical 
results for the vertical dots  can be compared to the experimental curves 
of Sasaki {\it et al.}\cite{sasaki}
The evolution of the differential conductance on the singlet side 
of the transition and at the crossover point shows similarity 
with the experiments, but the data of Sasaki {\em et al.} 
do not display the three-peak structure expected on the triplet side. 
The reason for this is that, the experiments were  performed 
at a finite temperature: the Kondo temperatures $T_{K}^{\pm}$
on the triplet side sink very quickly below the temperature as one 
gets farther away from the S-T degeneracy point, and therefore 
the  central Kondo resonance was unobservable.

\section{Conclusions}\label{conclu}

In this paper, we studied the non-equilibrium singlet triplet transition 
in lateral and in vertical quantum dots. We described the quantum
dots in terms of a  simple two-level Anderson model, which 
included the Hund's rule coupling ($J$) as well as the 
charging energy of the dot ($U$), and we used non-equilibrium 
iterative perturbation theory to
describe the transport through the dot.

As a first step, we explored the range of validity of perturbation
theory and found that it breaks down at intermediate coupling strengths, where 
multiple solutions and hysteresis appear as an indication of
spontaneous orbital polarization formation. 
 Then we computed the equilibrium
spectral functions, determined the non-equilibrium Green's functions,
and finally  used the Meir-Wingreen formula to compute the differential
conductance through the dot.

To our great
surprise, within its range of validity, this simple approach is able 
to describe essentially \emph{all 
experimentally-observed properties} of the transition. 
In particular,
for a lateral dot with level-dependent decay rates, $\Gamma_\pm$, the
differential conductance, $G(V)$, displays  
the correct resonance-antiresonance structure corresponding 
to the two-stage Kondo effect on the triplet side, 
while in the vicinity of the S-T transition 
these two resonances 
merge with the singlet excitation into a single Kondo resonance. 
On the singlet side of the
transition we find a split Kondo  resonance, and we recover the  correct
location of the side peaks, $eV\approx\pm (\Delta-2J)$.

For a lateral dot, the linear conductance, $G_{\rm lin}(\Delta)$ exhibits a broad maximum 
of  height $G_{\rm max}=2e^2/h$ around the transition, which is
located at the correct level splitting, $\Delta\approx 2J$.  
 On the other hand, for a vertical dot we find a maximal linear conductance of 
$4e^2/h$ on the triplet side of the transition, which gradually
crosses over to a small conductance on the singlet side. 
 We also verified that our numerical
calculations satisfy the Friedel sum rule, and that 
the linear conductance 
as computed by numerical differentiation of 
 the Meir-Wingreen formula is identical to the one
 computed from the phase shifts using the Landauer-B\"uttiker 
formula (assuming a Fermi liquid ground state).

The differential conductance of a vertical dot also behaves in the
correct way: A Kondo resonance (double resonance) is found on the 
triplet side, which is suppressed and broadened around the transition,
and finally splits into two resonances on the singlet side.  
  
We thus arrive at the remarkable and surprising conclusion that 
the perturbative approach is able to describe this rather 
complex behavior, and is in rather good agreement with 
the experiments.  It has however, some important limitations: 
(1) Its range of validity is limited to small and intermediate couplings, 
and (2) it is unable to account for the correct width of the Kondo
resonance. Furthermore, though we checked that our solutions satisfy current 
conservation, simple perturbation theory is not a conserving
approximation, and in general a lot of care must be taken to satisfy it.
However, as we show in a consecutive paper,\cite{future} the 
 problem of the width of the Kondo resonance as well as that of
 current conservation 
can be overcome at the expense of getting somewhat poorer resolution of
 the Hubbard peaks
by using the method of fluctuation exchange
approximation.\cite{future,white}

Acknowledgement: This research has been supported by Hungarian grants 
OTKA No. K73361 and OTKA NN76727,
the EU GEOMDISS project and its NKTH supplementary funding,
OMFB-00292/2010, and the Romanian grant CNCSIS PN II ID-672/2009.

\appendix
\section{Hybridized non-interacting Green's functions}\label{firstapp}

For completeness, here we enumerate 
 the non-interacting Green's functions, $\mathbf{g}$. 
They can be easily constructed based on the Dyson equation  
\begin{equation}
  \mathbf{g}^{-1}(\omega)
=\mathbf{g}_{0}^{-1}(\omega)-\mathbf{\Sigma}_{\Gamma}(\omega)\;, 
\label{eq:dyson2}
\end{equation}
where we used the matrix notation of the main text. 
The Green's functions, $\mathbf{g}_{0}$, in 
Eq.~\eqref{eq:dyson2} denote
 the Green's functions of the isolated  dot, with energies replaced by
 the  effective energies, $\tilde{\varepsilon}_{i\sigma}$, 
and its matrix elements are given by 
\begin{equation}
  g_{0\;i\sigma,i'\sigma'}^{\;\kappa\kappa'\;-1}=s_{\kappa}\; \delta_{ii'}\delta_{\sigma\sigma'}\delta_{\kappa\kappa'}\;
(\omega-\tilde{\varepsilon}_{i\sigma})\;,
\end{equation}
with $s_{\kappa}$ the Keldysh sign defined in the main text. 
The self-energy,
$\mathbf{\Sigma}_{\Gamma}$, denotes the self-energy coming from 
the hybridization of the dot and the leads, and its matrix elements
can be computed in terms of the tunneling matrix elements, $t_{i\alpha}$,
\begin{eqnarray}
  \Sigma_{\Gamma\;i\sigma,i'\sigma'}^{T} & = &
  i\;\pi\;\delta_{\sigma\sigma'}\sum_{\alpha=L,R}
  t_{i\alpha}t_{i'\alpha}
(2f_{\alpha}(\omega)-1)\;,
\nonumber
\\ 
  \Sigma_{\Gamma\;i\sigma,i'\sigma'}^{<} & = &
  -i\;2\pi\;\delta_{\sigma\sigma'}\sum_{\alpha=L,R}
 t_{i\alpha}t_{i'\alpha}\;f_{\alpha}(\omega)
\; ,\nonumber\\ 
  \Sigma_{\Gamma\;i\sigma,i'\sigma'}^{>} & = &
  -i\;2\pi\; \delta_{\sigma\sigma'}
\sum_{\alpha=L,R} t_{i\alpha}t_{i'\alpha}\;(f_{\alpha}(\omega)-1)\;,\nonumber\\ 
  \Sigma_{\Gamma\;i\sigma,i'\sigma'}^{\tilde{T}} & = &
  i\;\pi\;\delta_{\sigma\sigma'}\sum_{\alpha=L,R}t_{i\alpha}t_{i'\alpha}\;(2f_{\alpha}(\omega)-1)
  \;, \nonumber
\end{eqnarray}
with $f_{\alpha}(\omega)$ denoting the chemical potential-shifted 
Fermi functions of the leads.
In the numerical calculations, we inverted numerically 
 Eq.~\eqref{eq:dyson2} to obtain $\mathbf{g}$.

\end{document}